%% file: Template.tex
\documentclass{article}
\usepackage{spconf}

\usepackage{cite}
\usepackage{amsmath,amssymb,amsfonts}
\usepackage{algorithmic}
\usepackage{graphicx}
\usepackage{textcomp}
\usepackage{xcolor}

\usepackage{hyperref}
\hypersetup{
    colorlinks=true,
    citecolor=blue,
}
\usepackage{colortbl}

\usepackage{cite}
\usepackage{xurl}
\usepackage{amsfonts}
\usepackage{multirow}
\usepackage{makecell}
\usepackage{booktabs} % toprule, cmidrule, ...
\usepackage{subcaption}

\title{The VoiceMOS Challenge 2026: Evaluating Speech Enhancement, Emotional TTS and Accented TTS Systems}
\name{
    \begin{tabular}{c}
        Wen-Chin Huang$^{1}$,
        Wei Wang$^{2}$,
        Marvin Sach$^{3}$,
        Xiaoxue Gao$^{4}$, \\
        Nicholas Sanders$^{5}$,
        Erica Cooper$^{6}$,
        Tomoki Toda$^{1}$
    \end{tabular}
}
\address{
    $^{1}$Nagoya University, Japan
    $^{2}$Shanghai Jiao Tong University, China \\
    $^{3}$Technische Universität Braunschweig, Germany \\
    $^{4}$The Chinese University of Hong Kong, Shenzhen, China
    $^{5}$University of Edinburgh, UK \\
    $^{6}$National Institute of Information and Communications Technology, Japan\\
}
\begin{document}
\ninept
\maketitle
\begin{abstract}
We present the results of the VoiceMOS Challenge 2026, the fifth edition of a scientific challenge on automatic prediction of subjective speech assessments. After expanding the scope to music and general audio in 2025, we refocused the evaluation target on speech and organized three tracks: prediction of absolute and comparative category ratings for enhanced speech, prediction of naturalness and emotion-related tasks for emotional text-to-speech systems, and prediction of speaker and accent similarity for codec-based speech synthesis systems. The challenge attracted a total of 18 teams worldwide, with most teams successfully surpassing the provided baselines. We summarize the challenge results, representative top-performing systems, participant feedback, and directions for future editions.

\end{abstract}

\section{Introduction}
\label{sec:intro}

Founded in 2022, the VoiceMOS Challenge (VMC) series \cite{vmc2022, vmc2023, vmc2024, amc2025} aims to use standardized datasets in diverse and challenging domains to understand and compare techniques for predicting human ratings of speech, specifically those collected through mean opinion score (MOS) tests, hence the name VoiceMOS Challenge. The main goal has been to foster the development of automatic, data-driven speech quality assessment approaches to overcome the costly and time-consuming human listening tests that are conventionally regarded as the gold standard for evaluating speech.

Over the past four editions, the challenge has expanded its scope to reflect technological developments and feedback from the research community. Over the years, challenge tracks have ranged from in-domain to zero-shot settings, from English to multilingual settings, from single-axis to multidimensional evaluation targets, and from text-to-speech (TTS) to noisy and enhanced speech, and even singing voices. In 2025, we took a large step further and organized the AudioMOS Challenge (AMC) \cite{amc2025}, where we expanded the scope to music and general synthetic audio. Despite its success, we received feedback from the community that the main problems in speech evaluation remain unsolved. We thus decided to return to VMC and refocus on speech.

The VoiceMOS Challenge 2026\footnote{\url{https://sites.google.com/view/voicemos-challenge/voicemos-challenge-2026}} consists of three tracks. The first track focused on absolute category rating (ACR) and comparative category rating (CCR) prediction for speech generated by six systems from the URGENT 2026 speech enhancement challenge \cite{urgent2026}. The second track aimed to predict MOS for speech naturalness and emotion similarity in emotional TTS systems, with optional subtasks for predicting emotion-related labels. The third track was based on the CodecMOS-Accent dataset \cite{codecmos-accent}, and the task was to predict MOS for speaker and accent similarity in codec-based speech synthesis systems. In total, \textbf{18} unique teams submitted their final predictions, with 4/8/7 submissions for the three tracks, respectively. As a main result of the challenge, we are pleased to find that the baseline systems in each track were outperformed by most participating teams, demonstrating substantial progress over the challenge baselines.
Questionnaires submitted by participants also helped us identify successful techniques used by top-performing systems, as well as future directions suggested directly by the research community.

\input{tables/tracks}

\section{Challenge Description}
\label{sec:description}

The challenge took place from May 20th to August 10th, 2026.  The evaluation set speech samples for all tracks were released to participants on July 31st.  Predictions were due on August 10th, and the results were released to the participants on August 31st.
The challenge was held on CodaBench\footnote{\url{https://www.codabench.org/competitions/16419/}}
Each team was asked to submit a questionnaire to describe their system and provide feedback.  Table \ref{tab:datasets} summarizes the datasets used in each track. The details are described next. 

\subsection{Track 1: Predicting ACR and CCR of speech quality for speech enhancement systems
}

The Track 1 data \cite{urgent2026} were derived from the URGENT 2026 speech enhancement challenge and contain outputs from six participating systems for 840 source utterances across nine languages. The source speech includes both real recorded and simulated noisy conditions. For each utterance, the dataset contains six enhanced samples for ACR and all 15 pairwise system comparisons for CCR. In the ACR test, listeners rated overall speech quality on a five-point scale to obtain the MOS. In the CCR test, listeners compared two enhanced samples on a scale from $-3$ to $+3$ to obtain a comparative mean opinion score (CMOS), where a positive value indicates a preference for the first sample. Each sample or pair typically received eight ratings. Participants were required to predict both MOS and CMOS. The development and test sets were disjoint in both speakers and source utterances. The development set comprises 168 source utterances, 1,008 ACR samples, and 2,520 CCR pairs, while the test set comprises 672 source utterances, 4,032 ACR samples, and 10,080 CCR pairs.

\subsection{Track 2: Predicting MOS of speech naturalness and emotion similarity for emotional TTS systems}

The Track 2 data \cite{cooper2026interspeech} consists of emotional TTS samples in five different emotion categories (neutral, happy, sad, angry, surprised) and generated by 13 different synthesis systems, with sentences and natural reference audio coming from the ESD \cite{zhou2022emotional} and DailyTalk \cite{lee2023dailytalk} datasets.  Between 4 and 9 listeners rated each sample, and each listener rated the following factors: QMOS, an assessment of the synthesis quality based on pronunciation, intonation,
and signal quality; EMOS, an assessment of how well the speech matches the ground-truth emotion label; a categorical choice question where listeners select which of the emotion categories they can perceive in the audio; as well as ratings for valence, arousal, and dominance (V/A/D).  All ratings except for the
categorical emotion choices are on a five-point scale, and for the categorical question, listeners could choose more than one category.  For the challenge, QMOS and EMOS were required to predict, and the other questions were optional.

\subsection{Track 3: Predicting MOS of speaker and accent similarity for codec-based speech synthesis systems}

Track 3 targets accented English speech generated by codec-based speech synthesis systems. This track is based on the CodecMOS-Accent dataset \cite{codecmos-accent}, which in total contains 4,000 samples from 24 contemporary codec resynthesis and TTS systems, featuring 32 speakers across ten distinct accents. Given an input speech sample and a reference speech sample, the system outputs the speaker (SPK) and the accent (ACC) similarity score, or both. The dev and test sets are designed to contain samples from systems unseen during training.

\input{tables/teams}
\input{figs/main-results}

\section{Participants and baseline systems}
\label{sec:participants}

Table~\ref{tab:teams} shows the participants, their affiliations, and the tracks in which they participated.
In total, we received evaluation phase submissions from 18 unique teams spanning from countries and regions worldwide, with 15 teams from academia, 2 from industry, and 1 unknown.
Participants span from Germany, Japan, South Korea, Taiwan, France, China, India, USA, Brazil, Spain, and Canada.
For each track, we had 4, 8, and 7 participating teams, respectively,  with only one team participating in two tracks.
We individually informed each team of their randomly-assigned team ID.

We prepared open-sourced baseline systems for each track\footnote{\url{https://github.com/voicemos-challenge/vmc2026-baselines}}. 
The baseline system for track 1 was the UrgentMOS system \cite{urgentmos}, noted as B01. It is a unified non-intrusive speech quality assessment model supporting both absolute and comparative quality prediction. We used the publicly released \texttt{urgent-mos-f1c1m5dcorpus} checkpoint\footnote{\url{https://huggingface.co/urgent-challenge/urgent-mos-f1c1m5dcorpus}}, which employs WavLM \cite{wavlm} as its feature encoder and is trained on multiple speech-quality corpora. For ACR, the model predicts MOS from a single speech sample. For CCR, its preference module directly predicts CMOS from a pair of samples. All baseline predictions were obtained without fine-tuning on the challenge data.

The baseline system for track 2, noted as B02, consists of a different predictor for each question type asked to the listeners.  For QMOS, the predictor was UTokyo SaruLab UTMOS \cite{utmos}.  For EMOS, valence, dominance, and arousal, the predictor was Gemini LLM-as-judge using the \texttt{gemini-3-flash-preview}\footnote{\url{https://ai.google.dev/gemini-api/docs/models/gemini-3-flash-preview}} model prompted with the same instructions given to human listeners in the listening test.  For the emotion categories, the predictor was the category prediction probabilities from Emotion2vec+ large \cite{ma2024emotion2vec}.  Baseline predictions are obtained in a zero-shot manner without any finetuning.

There are two baseline systems for track 3. The first baseline, noted as B03-1, is a zero-shot system that simply outputs the cosine similarity of the embeddings of the two samples using a pre-trained ECAPA-TDNN \cite{ecapa-tdnn} speaker embedding model\footnote{\url{https://huggingface.co/speechbrain/spkrec-ecapa-voxceleb}} for both speaker and accent similarity scores. The second baseline, noted as B03-2, fine-tunes the pre-trained ECAPA-TDNN model using the training set. Two separate models were trained with the speaker and accent similarity scores, both with a batch size of 16, the AdamW optimizer with learning rate 0.001, and a fixed number of training steps of 20,000.

\section{Results}
\label{sec:results}

% \subsection{Evaluation metrics}

% \textcolor{red}{Except for those special subtracks in Track 2,} the system-level (SYS-) and utterance-level (UTT-) mean squared error (MSE), linear correlation coefficient (LCC), Spearman rank correlation coefficient (SRCC), and Kendall's Tau rank correlation coefficient (KTAU) were used as the evaluation metrics. The primary metric that is used to decided the final rankings of the challenge is UTT-SRCC, different from the SYS-SRCC adopted in previous challenges, as feedback from participants suggested that SYS-SRCC is often saturated and difficult to improve on.

% Due to space limits, in this paper, we will only show figures related to the main metrics, and the raw scores and the rankings of each team will be shown on the challenge website. 

% \input{figs/track1_scatter}
% \input{figs/track2_scatter}
% \input{figs/track3_scatter}
\input{figs/track1-analysis}

\subsection{Track 1 results}

For track 1, the primary metrics were the utterance-level Spearman rank correlation coefficient (UTT-SRCC) for the ACRs (ACR-UTT-SRCC) and the CCRs (CCR-UTT-SRCC). Figure~\ref{fig:track1-srcc} shows the scatter plot of the results from each participant. All participants were able to outperform the baseline, with T09, T01 being the top performing teams for ACR-UTT-SRCC and CCR-UTT-SRCC, respectively. Aside from the individual results, we observed that \textit{CCR was more difficult to predict than ACR}, as the top system achieved an ACR-UTT-SRCC of 0.779, whereas the top CCR-UTT-SRCC was 0.487.
An additional observation was made in Figure~\ref{fig:track1-analysis}, which shows that CCR prediction becomes easier as the magnitude of the ground-truth CCR increases. In particular, the sign accuracy consistently increases with \(|\mathrm{CCR}|\), indicating that systems are more likely to predict the correct preference direction when the perceptual difference between the two samples is larger.

\subsection{Track 2 results}

For track 2, the primary metrics were UTT-SRCC for QMOS (QMOS-UTT-SRCC) and EMOS (EMOS-UTT-SRCC). We also calculated UTT-SRCC for V/A/D, and for the categorical prediction we defined a metric, CAT-ERR, where we evaluated the categorical predictions' match to the human category choices by taking the absolute value of the difference between the predicted value for each category and the percent of human listeners choosing that category for the given audio sample.  Then, the averaged error value over all emotion categories is computed.   %% we can make this shorter if we don't report any results on it

\input{figs/track2-vad-cat}

It can be observed in Figure \ref{fig:track2-srcc} that for QMOS and EMOS, most of the teams were able to outperform the baseline. \textit{EMOS was not necessarily more difficult to predict than QMOS}, with the top systems achieving UTT-SRCCs of 0.785 and 0.758, respectively.
All teams except for two (T13 and T18) participated in the optional sub-tasks (category choices and V/A/D) and all of those teams also solidly beat the baseline on utterance-level metrics (Figure \ref{fig:track2-vad-cat}).  Out of the three emotional dimensions of V/A/D, arousal (A) was consistently the easiest one to predict.

Considering whether some emotion categories might be easier or more difficult to predict, we surprisingly found that the Neutral category was slightly more difficult for predicting QMOS and EMOS for most of the teams, while predicting QMOS for the Sad category tended to be slightly easier.  The Happy and Sad categories were easiest for predicting valence, and the Angry and Sad categories were easiest for predicting arousal and dominance.  For the emotion categories task, Neutral tended to be over-predicted for most teams, whereas Surprise tended to be under-predicted.

\subsection{Track 3 results}

For track 3, the primary metrics were the utterance-level Spearman rank correlation coefficient for the speaker similarity (SPK-UTT-SRCC) and the accent similarity (ACC-UTT-SRCC). Figure~\ref{fig:track3-srcc} shows the scatter plot of the results from each participant. Most participants were able to outperform both baselines, with T04 being the top performing team as they ranked first on both SPK-UTT-SRCC and ACC-UTT-SRCC. We also observed that most teams achieved higher SPK-UTT-SRCC than ACC-UTT-SRCC, indicating that speaker similarity was easier to predict than accent similarity.

\section{Insights from the questionnaires}

\subsection{Brief descriptions of the top performing systems}
\label{ssec:top-systems}

% \subsubsection{\textbf{T01} and \textbf{T09}, top performing systems in track 1}

\textbf{T01} and \textbf{T09} were the \textbf{two top systems in CCR-UTT-SRCC and ACR-UTT-SRCC in Track 1}. The T01 system, trained on BVCC \cite{bvcc}, NISQA \cite{nisqa}, SingMOS-Pro \cite{singmos-pro}, TMHINT-QI \cite{tmhintqi}, URGENT2024-SQA \cite{urgent2024}, and URGENT2025-SQA \cite{urgent2025}, fused frozen wav2vec 2.0 large robust \cite{wav2vec2} and MERT \cite{mert} features to predict frame-level quality scores, which virtual listeners pooled into utterance-level candidates. The model jointly learned the mean and inter-listener standard deviation of utterance scores, with a listener router producing the final ACR prediction. Training also used partial mix-up augmentation, supervised contrastive learning, and an utterance-ranking loss. The CCR was simply the lipped difference of the predicted ACRs.

The T09 system predicts ACR with a weighted ensemble of 29 scores, including 27 from several pre-trained SQA models and two dual-branch  models obtained by fine-tuning Whisper medium \cite{whisper} and HuBERT large \cite{hubert} encoders with NISQA, TMHINT-QI, and URGENT 2025. Separate ensemble weights for ACR and CCR were fitted on the Track 1 dev set labels. For CCR, 29 score differences were rank-normalized and combined by the CCR ensemble weights.

% \subsubsection{\textbf{T02}, top performing system in track 2}

\textbf{T02 ranked first in QMOS-UTT-SRCC, EMOS-UTT-SRCC, UTT-SRCC for valence and arousal, as well as CAT-ERR in Track 2}. Their QMOS model combined a listener embedding \cite{ldnet}, frozen UTMOS features \cite{utmos}, and a WavLM Large encoder \cite{wavlm} fine-tuned to predict QMOS and emotion labels. The EMOS model combined a listener embedding and WavLM Base Plus and Large encoders fine-tuned with emotion labels only. Predictions from both models were averaged across 64 listener identities and a set of cross-validated models.

% \subsubsection{\textbf{T04}, top performing system in track 3}

\textbf{T04 ranked first in both SPK-UTT-SRCC and ACC-UTT-SRCC in Track 3}.
Separate systems were built to predict SPK and ACC individually, both of which ensembled eight members that varied in their frozen pretrained speech embeddings, reference representations, loss functions, and checkpoint-selection criteria. The official training data was used, with VoxSim \cite{voxsim} additionally used to warm-start most SPK models.
Models were trained on individual ratings conditioned on learned embeddings for the 25 listeners. At inference, predictions were averaged across all listener embeddings and calibrated ensemble members. Additionally, SPK predictions underwent a system-level de-shrinkage correction process.

\subsection{Feedback from the participants}

Most participant feedback was positive, praising the timely and relevant design of the tracks, the convenience of the submission platform, and the organizers’ responsiveness. Negative comments included the lack of advance notice about the release and permitted use of development-set labels, limited or imbalanced dataset sizes, and insufficient diagnostic information for failed submissions. Suggestions about the future tasks include (1) explainable and uncertainty-aware SQA, (2) evaluation for long-form speech or full-duplex spoken dialogues, (3) cross-lingual/-domain evaluation and (4) automatic prediction of outcomes from preference tests.

\section{Conclusion}

The VoiceMOS Challenge 2026 covered three different speech quality assessment settings, and most participating teams were able to outperform the provided baselines. Beyond the leaderboard results, several observations emerged from the individual tracks. In track 1, CCR prediction appeared to be more difficult than ACR prediction. In track 2, EMOS was not necessarily more difficult to predict than QMOS, while in track 3, accent similarity was slightly more difficult to predict than speaker similarity. The top-performing submissions also often relied on relatively complex combinations of pretrained models, listener modeling, and ensembling. While their effectiveness is evident from the challenge results, the trade-offs between such system complexity and properties such as interpretability, robustness, and generalization remain an interesting direction for future investigation.

Participant feedback also pointed to several directions for future challenges, including explainable SQA, evaluation for long-form speech and full-duplex spoken dialogues, cross-lingual and cross-domain evaluation, and automatic prediction of outcomes from preference tests. We also surveyed participants about the availability of large-scale listening-test data that could potentially be shared for future editions. Several participants indicated interest in contributing such resources, and we have begun exploring follow-up data collection efforts based on these responses.

\section{ACKNOWLEDGMENTS}
\label{sec:ack}

% \noindent\textbf{Acknowledgements}:
This work was partly supported by JSPS KAKENHI Grant Number 25K00143.

\bibliographystyle{IEEEtran}
\bibliography{strings,refs}

\end{document}

%% file: tables/tracks.tex
\begin{table*}[t]
	\centering
	\caption{Summary of each track.}
	
	\centering
	\begin{tabular}{ c c c c c c c c c c}
		\toprule
		\multirow{2}{*}[-3pt]{Track} & \multirow{2}{*}[-3pt]{\begin{tabular}[l]{@{}c@{}}Evaluated\\systems\end{tabular}} & \multirow{2}{*}[-3pt]{\begin{tabular}[l]{@{}c@{}}Prediction\\targets\end{tabular}} & \multicolumn{3}{c}{\# Samples} & \multicolumn{3}{c}{\# Systems} & \multirow{2}{*}[-3pt]{\makecell{\# ratings\\per sample}}\\
		\cmidrule(lr){4-6} \cmidrule(lr){7-9}
		& & & Train & Dev & Test & Train & Dev & Test & \\
		\midrule
		1 & \begin{tabular}[c]{@{}c@{}}Speech\\enhancement\end{tabular} & ACR, CCR & -- & \begin{tabular}[c]{@{}c@{}} ACR: 1008\\CCR: 2520\end{tabular} & \begin{tabular}[c]{@{}c@{}}ACR: 4032\\CCR:10080\end{tabular} & -- & 6 & 6 & 8 \\
		\midrule
		2 & \begin{tabular}[c]{@{}c@{}}Emotional\\TTS\end{tabular} & \begin{tabular}[c]{@{}c@{}}QMOS, EMOS\\optional: V/A/D,\\perceived emotion\end{tabular} & 12746 & 2730 & 2730 & 10 & 13 & 14 & 4-9\\
            \midrule
            3 & \begin{tabular}[c]{@{}c@{}}Accented\\TTS\end{tabular} & \begin{tabular}[c]{@{}c@{}}Speaker \& accent\\similarity\end{tabular} & 2800 & 600 & 600 & 21 & 23 & 25 & 4.9\\
		\bottomrule
	\end{tabular}
	\label{tab:datasets}
    \vspace{-3mm}
\end{table*}

% Set name	Number of ACR samples	Number of CCR samples
% Validation set	1008	2520
% Evaluation set	4032	10080

%% file: tables/teams.tex
\begin{table}[t]
    % \footnotesize

    \centering
    \caption{List of participant affiliations, in random order.}
    \label{tab:teams}

    \begin{tabular}{@{}l|ccc@{}}
        \toprule
        \multicolumn{1}{c|}{\multirow{2}{*}{\textbf{Affiliation}}} & \multicolumn{3}{c}{\textbf{Track}}   \\ \cmidrule(l){2-4}
        & \textbf{1} & \textbf{2} & \textbf{3} \\ \midrule
        \begin{tabular}[l]{@{}l@{}}Advanced Knowledge Center for\\\hspace{0.5cm}Immersive Technologies, Brazil\end{tabular} &     V      &            &            \\
        Microsoft, Germany &     V      &            &            \\
        BITS Pilani, Hyderabad Campus, India &     V      &            &            \\
        Paderborn University, Germany &     V      &            &            \\
        UTokyo-AIST-KAIST, Japan \& South Korea &            &     V      &            \\
        Chinese Academy of Science, China &            &     V      &            \\
        Le Mans Université, France &            &     V      &            \\
        \begin{tabular}[l]{@{}l@{}}Friedrich-Alexander University of \\ \hspace{0.5cm} Erlangen–Nuremberg, Germany\end{tabular} &            &     V      &            \\
        Universidade Estadual de Campinas, Brazil &            &     V      &     V      \\
        California State University, USA$^\ddagger$ &            &     V      &            \\
        Ningbo University, China$^{\dagger}$  &            &     V      &            \\
        Universitat Pompeu Fabra, Spain$^\ddagger$ &            &     V      &            \\
        Soul App, China &            &            &     V      \\
        Hankuk University of Foreign Studies, South Korea &            &            &     V      \\
        National Taiwan University &            &            &     V      \\
        Individual, India &            &            &     V      \\
        Mila-NRC, Canada &            &            &     V      \\
        Ningbo University, China$^{\dagger}$$^\ddagger$ &            &           &  V          \\
        \bottomrule
        \multicolumn{4}{l}{
            {\footnotesize $^\dagger$: These teams self-claimed to be independent from each other.}
        }\\
        \multicolumn{4}{l}{
            {\footnotesize$^\ddagger$: Did not submit the questionnaire.}
        } \\
    \end{tabular}
    \vspace{-4mm}
\end{table}

%% file: figs/main-results.tex
\begin{figure*}[t]
    \centering

    \begin{subfigure}[t]{0.33\textwidth}
        \centering
        \includegraphics[width=\linewidth]{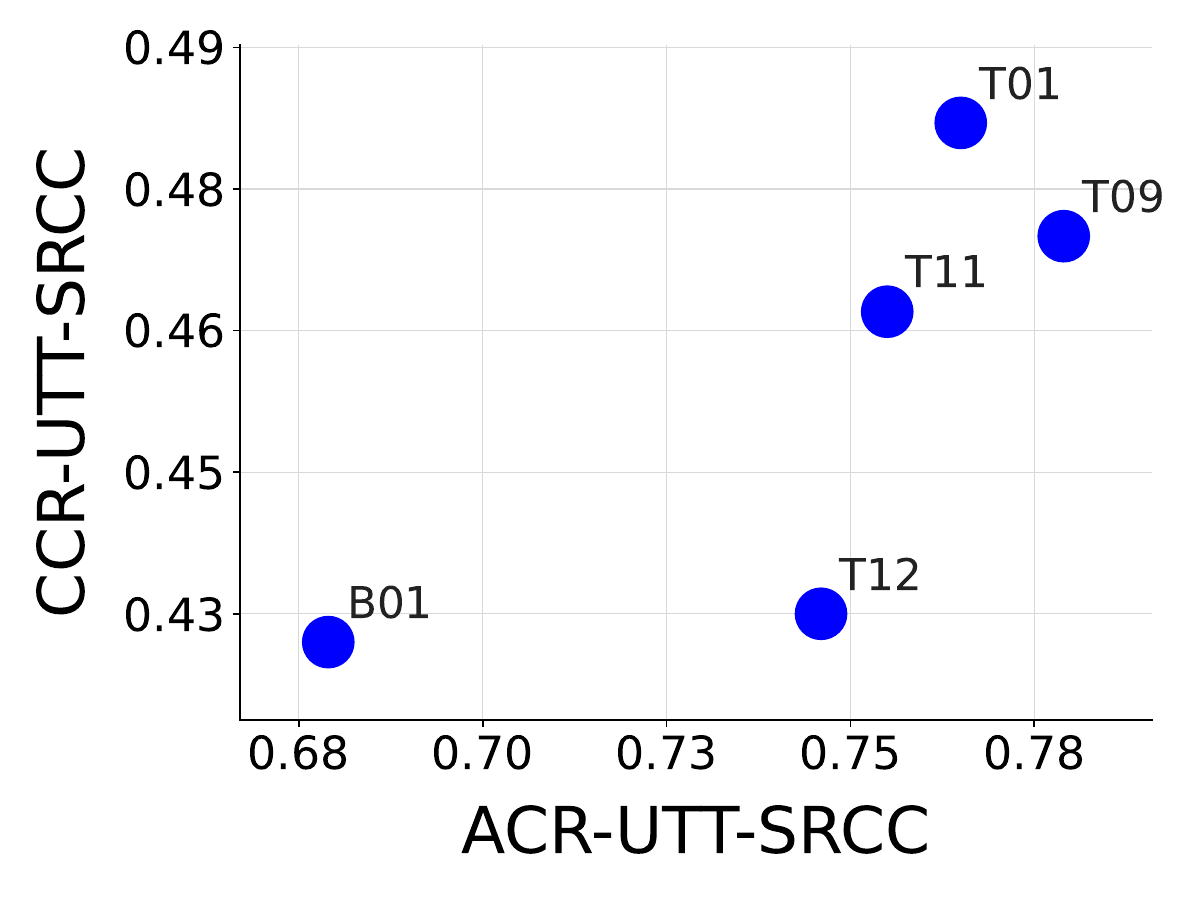}
        \caption{Track 1}
        \label{fig:track1-srcc}
    \end{subfigure}
    \hfill
    \begin{subfigure}[t]{0.33\textwidth}
        \centering
        \includegraphics[width=\linewidth]{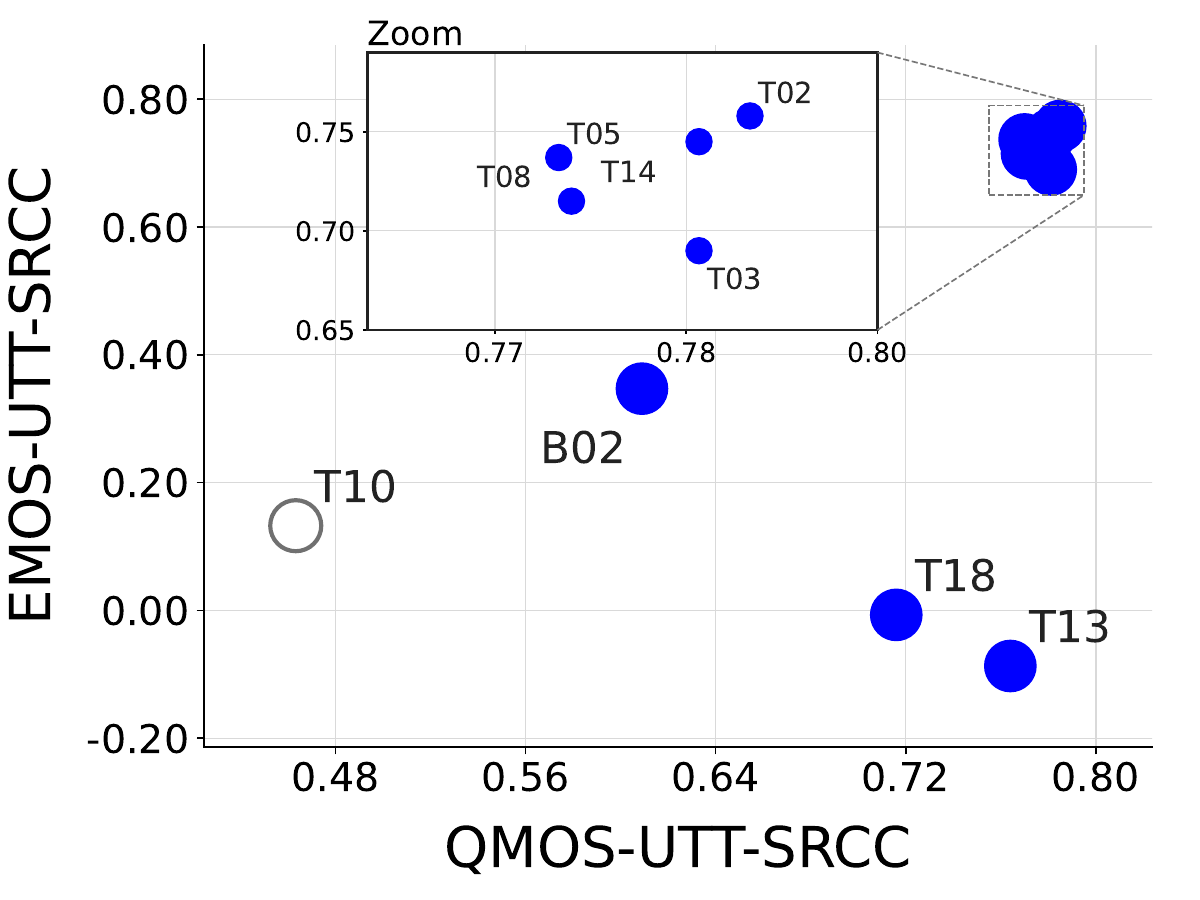}
        \caption{Track 2.}
        \label{fig:track2-srcc}
    \end{subfigure}
    \hfill
    \begin{subfigure}[t]{0.33\textwidth}
        \centering
        \includegraphics[width=\linewidth]{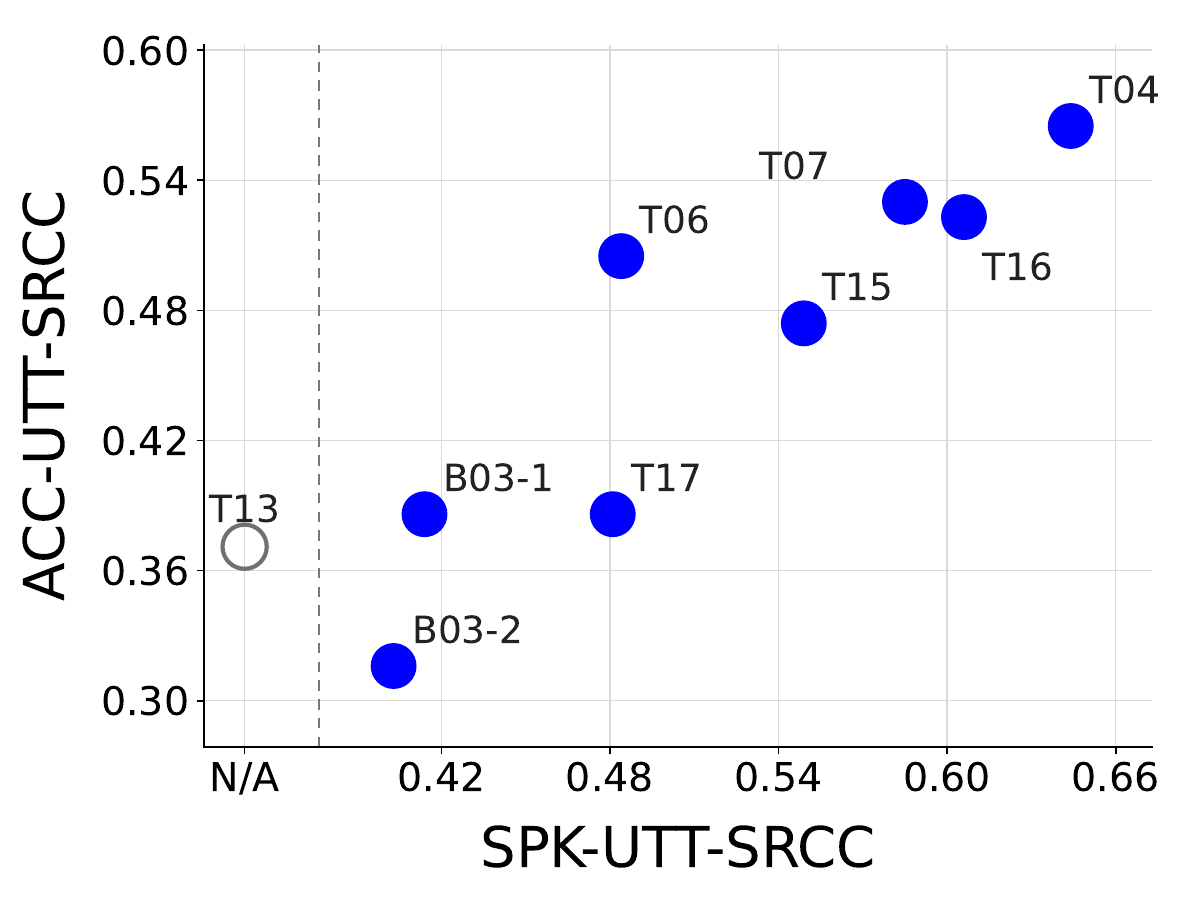}
        \caption{Track 3.}
        \label{fig:track3-srcc}
    \end{subfigure}

    \caption{Utterance-level SRCC results for the three tracks.
    Hollow gray markers denote provisional or incomplete submissions.  There were 157 missing samples in the T10 submission, and metrics were calculated using the 2574 available samples. T13 only submitted ACC results.}
    \label{fig:track-srcc-comparison}
    \vspace{-5mm}
\end{figure*}

%% file: figs/track1-analysis.tex
\begin{figure}[t]
	\centering
	\includegraphics[width=0.7\linewidth]{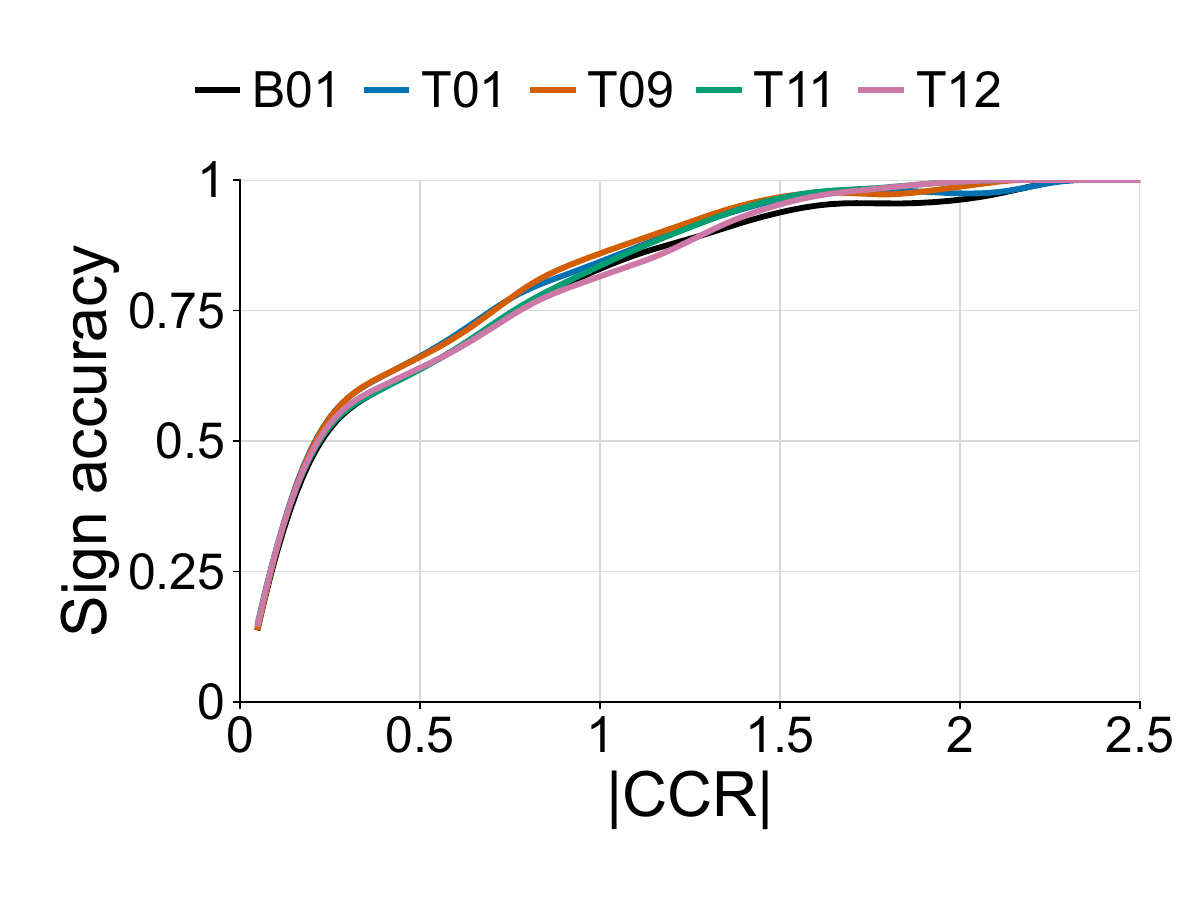}
    \vspace{-10pt}
	\caption{
        \label{fig:track1-analysis}
        Sign accuracy of Track 1 CCR predictions as a function of \(|\mathrm{CCR}|\), where sign accuracy measures whether the predicted CCR has the same sign as the ground-truth CCR. Results are computed over 0.1-wide \(|\mathrm{CCR}|\) bins and smoothed for visualization.
    }
\vspace{-4mm}
\end{figure}

%% file: figs/track2-vad-cat.tex
\begin{figure}[t]
	\centering
	\includegraphics[width=\linewidth]{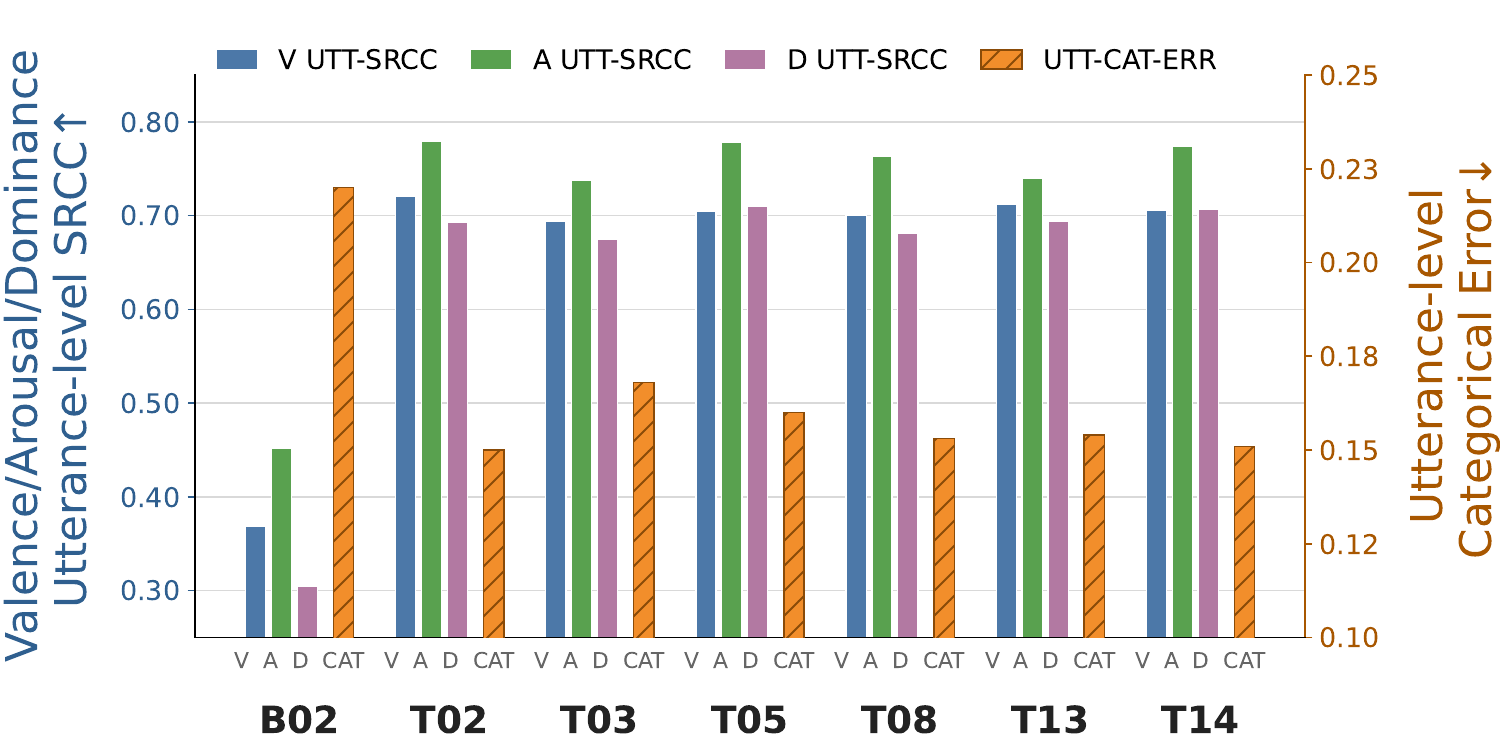}
    % \vspace{-10pt}
	\caption{
        \label{fig:track2-vad-cat}
        Track 2 optional emotion-related prediction task results. Bars show UTT-SRCC for valence (V), arousal (A), and dominance (D), together with UTT-CAT-ERR for emotion-category prediction. UTT-SRCC values are shown on the left axis, while UTT-CAT-ERR is shown on the right axis; higher UTT-SRCC and lower UTT-CAT-ERR indicate better performance.
    }
\vspace{-3mm}
\end{figure}